\documentclass{aa}

\usepackage{graphicx}
\usepackage[varg]{txfonts}
\usepackage{empheq}

\DeclareSymbolFont{txlargeoperators}{OMX}{txex}{m}{n}
\DeclareSymbolFont{txlargeoperatorsA}{U}{txexa}{m}{n}
\DeclareMathSymbol{\intop}{\mathop}{txlargeoperators}{"52}
\DeclareMathSymbol{\iintop}{\mathop}{txlargeoperatorsA}{33}
\DeclareMathSymbol{\iiintop}{\mathop}{txlargeoperatorsA}{35}
\DeclareMathSymbol{\iiiintop}{\mathop}{txlargeoperatorsA}{37}
\DeclareMathSymbol{\idotsintop}{\mathop}{txlargeoperatorsA}{39}

\DeclareMathSymbol{\ointop}{\mathop}{txlargeoperators}{"48}
\DeclareMathSymbol{\oiintop}{\mathop}{txlargeoperatorsA}{8}
\DeclareMathSymbol{\oiiintop}{\mathop}{txlargeoperatorsA}{41}

\usepackage{mathrsfs}  
\usepackage{color}
\usepackage{paralist}
\usepackage{enumitem}

\usepackage{scalerel}
\newcommand\wh[1]{\hstretch{2}{\hat{\hstretch{.5}{#1}}}}

\usepackage{stackengine}
\renewcommand\dot[1]{\stackMath\stackengine{1pt}{#1}{\mbox{\large\bfseries .}}{O}{c}{F}{T}{S}}
\renewcommand\ddot[1]{\stackMath\stackengine{1pt}{#1}{\mbox{\large\bfseries .\hspace{-0.1ex}.}}{O}{c}{F}{T}{S}}

\let\oldsqrt\sqrt
\renewcommand\sqrt[1]{\!{\textstyle\oldsqrt{#1}}\,}
\def\sqrt#1{\!\oldsqrt{#1}\,}
\usepackage{natbib,twoopt}
\usepackage[breaklinks=true]{hyperref} 
\hypersetup{colorlinks=true,linkcolor=blue,filecolor=blue,citecolor=blue,urlcolor=blue}
\bibpunct{(}{)}{;}{a}{}{,}             
\makeatletter
  \newcommandtwoopt{\citeads}[3][][]{\href{http://adsabs.harvard.edu/abs/#3}%
    {\def\hyper@linkstart##1##2{}%
     \let\hyper@linkend\@empty\citealp[#1][#2]{#3}}}
  \newcommandtwoopt{\citepads}[3][][]{\href{http://adsabs.harvard.edu/abs/#3}%
    {\def\hyper@linkstart##1##2{}%
     \let\hyper@linkend\@empty\citep[#1][#2]{#3}}}
  \newcommandtwoopt{\citetads}[3][][]{\href{http://adsabs.harvard.edu/abs/#3}%
    {\def\hyper@linkstart##1##2{}%
     \let\hyper@linkend\@empty\citet[#1][#2]{#3}}}
  \newcommandtwoopt{\citeyearads}[3][][]%
    {\href{http://adsabs.harvard.edu/abs/#3}
    {\def\hyper@linkstart##1##2{}%
     \let\hyper@linkend\@empty\citeyear[#1][#2]{#3}}}
\makeatother

\usepackage{orcidlink}
\usepackage{balance}

\begin{document}
\title{Fast particle-mesh code for Milgromian dynamics}
\author{P.M.\ Visser \inst{1} \and S.W.H.\ Eijt \inst{2} \and J.V.\ de Nijs \inst{1,2} \orcidlink{0009-0005-4827-2106}}
\institute{
Delft Institute of Applied Mathematics, Delft University of Technology,
Mekelweg 4, 2628 CD Delft, The Netherlands \\
\email{p.m.visser@tudelft.nl}
\and
Department of Radiation Science and Technology, Applied Sciences, Delft University of Technology,
Mekelweg 15, 2629 JB Delft, The Netherlands
}
\date{Received August 30, 2023\ / Accepted October 3, 2023}
\abstract
{Modified Newtonian dynamics (MOND) is a promising alternative to dark matter. To further test the theory, there is a need for fluid- and particle-dynamics simulations. The force in MOND is not a direct particle-particle interaction, but derives from a potential for which a nonlinear partial differential equation (PDE) needs to be solved. Normally, this makes the problem of simulating dynamical evolution computationally expensive.
}
{We intend to develop a fast particle-mesh (PM) code for MOND (the AQUAL formalism). 
}
{We transformed the nonlinear equation for MOND into a system of linear PDEs plus one algebraic equation. An iterative scheme with the fast Fourier transform (FFT) produces successively better numerical approximations.
}
{The algorithm was tested for dynamical systems in MOND where analytical solutions are known: the two-body problem, a body with a circular ring, and a spherical distribution of particles in thermal equilibrium in the self-consistent potential.
}
{The PM code can accurately calculate the forces at subpixel scale and reproduces the analytical solutions. Four iterations are required for the potential, but when the spatial steps are small compared to the kernel width, one iteration is suffices. The use of a smoothing kernel for the accelerations is inevitable in order to eliminate the self-gravity of the point particles. Our PDE solver is $15$ to $42$ times as slow as a standard Poisson solver. However, the smoothing and particle propagation takes up most of the time above one particle per $10^3$ pixels. The FFTs, the smoothing, and the propagation part in the code can all be parallelized.
}

\maketitle

\section{Introduction}
If stars and galaxies move under Newtonian gravity, dark matter is required to bind the stars to galaxies and galaxies in clusters. There is ample evidence. The mass inferred from the stellar velocities in clusters \citep[with the virial equation,][]{Zwicky1933,Zwicky1937} and in galaxies \citep[with Huygens' equation,][]{Rubin1970,Rubin1980} is much higher than the mass of the luminous matter. $\Lambda$CDM cosmology requires six times as much matter as can be directly detected, hence, a large nonradiative matter component must dominate the Universe.

Despite a search of over half a century, no dark matter particle has been directly detected, while the list of astronomical observations that are difficult to explain with dark matter models has been growing:
\begin{inparaenum}[(i)]
\item
The absence of cusps at galactic cores \citep{McGaugh2003,Blok2009};
\item
bar structures at galactic cores \citep{Chiba2021,Roshan2021,Roshan2021b};
\item
the low number of galaxy satellites \citep{Bullock2013};
\item
the absence of dynamical friction on galaxy satellites by its 
dark matter halo \citep{AngusKroupa2011,Kroupa2015,Oehm2017};
\item
the shape and central brightness of the Fornax cluster \citep{Asencio2022};
\item
the symmetry breaking of tidal tails of open star clusters \citep{Kroupa2022};
\item
galaxies without dark matter \citep{Moreno2022,Comeron2023};
\item
the intracluster gas does not cool \citep{Fabian1994};
\item
strong local lensing in galaxy clusters \citep{Meneghetti2020};
\item
the early formation of very large galaxies and clusters \citep{Asencio2020,Asencio2023,Labbe2023};
\item
the Hubble $H_0$ tension and the $S_8$ tension occur in standard $\Lambda$CDM cosmology \citep{Aghanim2020,Haslbauer2020,Asgari2021,Riess2021};
\item
the violation of the strong equivalence principle \citep{Blanchet2011,Chae2020};
\item
and the velocities in wide binary star systems 
\citep{Banik2018,Pittordis2019,Chae2023,Hernandez2023}.
\end{inparaenum}

An alternative to dark matter is modified Newtonian dynamics (MOND), which was invented by Milgrom and Bekenstein \citep{Milgrom1983,Bekenstein1984,Milgrom2008}. Whereas Einstein's general relativity modifies Newton when the gravity potential $\phi$ is deep, Milgrom's MOND instead modifies Newton at low accelerations. The critical acceleration scale is close to the scale set by the cosmological constant $\Lambda$. It is $a_0\approx .127\cdot c^2\Lambda^{1/2}$. The strengths of the mutual gravity between stars, the centripetal acceleration of stars in the galaxy outskirts, and the accelerations in galaxy clusters all have values of about $a_0$. Because MOND is nonrelativistic, it applies in the regime in which $|\phi|\ll c^2$ and $|\boldsymbol\nabla\phi|\lesssim c^2\Lambda^{1/2}$. The theory explains some of the listed problems and it gives a natural explanation for the observed flat rotation curves of galaxies as well as for the Tully-Fisher relation \citep{Mcgaugh2000}.

Although the MOND model makes definite predictions without the need for fit parameters, the model is complicated by the fact that it is not relativistic. It needs a preferred inertial reference frame (i.e.,\ it is an aether theory) and the gravity from distant sources cause an external field effect (violating the strong equivalence principle). These effects need to be considered when the possibility of MOND is to be ruled out or confirmed.

General relativistic models that reduce to MOND at low energies introduce additional degrees of freedom. The new TeVeS theories of \citet{Zlosnik2019,Zlosnik2021} have resolved the long-standing problem that gravitational waves traveled faster than photons. This problem plagued older MOND-extensions \citep{Bekenstein2004}.

Milgrom's MOND 
modifies the equation for the gravitational potential using a nonlinear PDE. The numerical codes that have been developed to simulate the dynamical evolution in MOND are found in \citet{BradaMilgrom1999,Nipoti2007,Llinares2008}, N-MODY by \citet{Londrillo2009}, and RayMOND by \citet{Candlish2014}, based on multigrid discretization. The work of \citet{Angus2011,Peng2015} solved for QuMOND, the quasilinear MOND model \citep{Milgrom2010}. The code called Phantom of Ramses \citep{Lughausen2015,Nagesh2022} also solves QuMOND. The codes RayMOND and Phantom of Ramses are both based on RAMSES \citep[by][]{Teyssier2002}.

We describe an algorithm for $N$-body simulations in MOND. The particles can be Solar System planets, stars in a galaxy, or the galaxies in a galaxy cluster. The following main fields in astronomy are affected by Milgromian gravity:

\begin{inparaenum}[(i)]
\item
Celestial mechanics. In the Solar System, the perturbation of the planetary orbits due to a possible residual MOND effect may have to be calculated. Because outer Solar System dynamics is in the crossover regime, this correction critically depends on the interpolation function, which describes the transition between MOND and normal gravity. In particular, the perihelion precession of the outer planets, where acceleration is lowest, could be affected \citep{Pitjeva2013,Pauco2016,Pauco2017,YoonDarriba2020}. Here, the effect of MOND may lie just below the current detection limit. Binary star systems with a wide separation are in the transition of the interpolation 
\citep{Hernandez2019}.
For both these systems a two-body model is appropriate when the external field from the Milky Way is taken into account \citep{Iorio2009,Iorio2017}.

\item
Dense stellar systems. These include open clusters and globular clusters, where close encounters between stars are relevant. The evolution requires inclusion of short-range gravitational interactions. Hence, individual particles must be tracked.

\item
Galactic dynamics. Examples are transient phenomena such as tidal effects between neighboring galaxies, the collisions between galaxies, the formation of tails, and the collapse or merging of smaller stellar systems. Galaxy-formation simulations in QuMOND were made by \citet{Wittenburg2020,Eappen2022}.

\item
Galaxy clusters. Here, the particles are the individual galaxies. An $N$-body code for the dynamics with small particle numbers, between $10^2$ and $10^3$, possibly supplemented with hydrodynamics for the gas component, would be the appropriate model.

\item
Cosmology. Structure formation in the early Universe would be affected by MOND because the density fluctuations will initially be small, accelerations could begin with low values, and structure formation may start in the deep-MOND regime. Cosmological simulations in QuMOND were made by \citet{Katz2013,Haslbauer2020,Wittenburg2023}.
\end{inparaenum}

\newpage
\section{Theory and main result: The idea of the method}
\label{sect2}
As the particles move in space under the influence of the gravitational potential $\phi(\boldsymbol r,t)$, they experience an acceleration given by Newton's second law,
\begin{equation}
m_i\ddot{\boldsymbol r}_i(t) = -m_i\boldsymbol\nabla\phi(\boldsymbol r_i(t),t)
.
\label{ddr}
\end{equation}
The baryonic mass density for the set of $N$ point masses is
\begin{equation}
\rho_\text{B}(\boldsymbol r,t) =  {\sum}_{i=1}^N m_i \delta(\boldsymbol r-\boldsymbol r_i(t))
\label{density}
.
\end{equation}
Here, $\delta$ is the three-dimensional Dirac-delta function. Whereas the potential is a solution of the Poisson equation in the Newton theory, in MOND (the AQUAL formalism), the potential is a solution to the following nonlinear partial differential equation (PDE):
\begin{equation}
\boldsymbol\nabla\bullet \mu\bigg(\frac{|\boldsymbol\nabla\phi|}{a_0}\bigg) \boldsymbol\nabla\phi = 4\pi G \rho_\mathrm{B}(\boldsymbol r,t)
,
\label{PDE}
\end{equation}
with the boundary condition $\boldsymbol\nabla\phi(\boldsymbol r,t) \longrightarrow \boldsymbol 0$ as $|\boldsymbol r| \longrightarrow \infty$. The scalar function $\mu(x)$ is called the interpolation function because it interpolates between the regime of high and low accelerations. At high acceleration, $\mu(x)\longrightarrow 1$ so that the potential will approach the Newton potential. At low acceleration, the deep MOND regime begins, and $\mu(x)\longrightarrow x$. An isolated and localized mass $m$ has the asymptotic spherically symmetric potential $\phi(\boldsymbol r)=\sqrt{Gma_0}\log|\boldsymbol r|$. 
See Table \ref{table1} 
for the symbols and notation. A popular choice for the interpolation function that is consistent with all observations so far \citep{Brouwer2021} is
\begin{equation}
\mu(x) = \frac{x}{\sqrt{1+x^2}}
.
\label{mu}
\end{equation}

\subsection{An equivalent system for AQUAL}
We now demonstrate how to rewrite the Eq.~(\ref{PDE}) for the potential in MOND by transforming this equation into a system of coupled linear PDEs for four vector fields $\boldsymbol g_\text{N}$, $\boldsymbol g_\text{M}$, $\boldsymbol H$, and $\boldsymbol F$, combined with one algebraic equation. This system of equations is
\begin{empheq}[left=\empheqlbrace]{align}
\boldsymbol\nabla \bullet \boldsymbol g_\text{N} &= -4\pi G \rho_\mathrm{B}
, &
\boldsymbol\nabla \times \boldsymbol g_\text{N} &= \boldsymbol 0
\label{newtongrav}
, \\
\boldsymbol F &= \boldsymbol g_\text{N} + \boldsymbol H
\label{helmholtz}
, & & \\
& & \boldsymbol\nabla \bullet \boldsymbol H &= 0
\label{divH}
, \\
\boldsymbol g_\text{M} &= \nu\bigg(\frac{F}{a_0}\bigg) \boldsymbol F
\label{inverse}
, & & \\
& & \boldsymbol\nabla \times \boldsymbol g_\text{M} &= \boldsymbol 0
\label{rotg}
, \\
\boldsymbol F &= \mu\bigg(\frac{g_\text{M}}{a_0}\bigg) \boldsymbol g_\text{M}
.
\label{interpolation}
\end{empheq}
The vector fields $\boldsymbol g_\text{N}$ and $\boldsymbol g_\text{M}$ are the Newton and MOND acceleration fields. With the identification $\boldsymbol g_\text{N}=-\boldsymbol\nabla\phi_\text{N}$, the two equations in Eq.~(\ref{newtongrav}) are equivalent to the Poisson equation. Our new system effectively expresses the MOND gravity field $\boldsymbol g_\text{M}$ in the Newtonian gravity field $\boldsymbol g_\text{N}$. Equations (\ref{inverse}) and (\ref{interpolation}) are not independent equations, but are the inverse of each other. Here, $F$ and $g_\text{M}$ are the lengths of the 
vectors $\boldsymbol F$ and $\boldsymbol g_\text{M}$. The pair $\mu(x)$ and $\nu(y)$ relate the strength of the Newtonian gravity to the strength of the Milgromian gravity by
\[
\mu(x)x = y = \frac{F}{a_0} , \quad \text{and} \quad \nu(y)y = x = \frac{g_\text{M}}{a_0}
.
\]
The choice Eq.~(\ref{mu}) for the interpolation function gives
\[
\nu(y) = \sqrt{\frac{1}{2}+\frac{1}{2}\sqrt{1+\frac{4}{y^2}}}
.
\]

Here follows the proof that the system of Eqs.~(\ref{newtongrav})-(\ref{interpolation}) is equivalent to Eq.~(\ref{PDE}). Instead of an equation for the potential $\phi$, we have an equation for the acceleration field $\boldsymbol g_\text{M}$. We thus need to identify $\boldsymbol g_\text{M}=-\boldsymbol\nabla\phi$. When we now combine Eq.~(\ref{helmholtz}) with (\ref{interpolation}), solve for $\boldsymbol g_\text{N}$, then substitute this solution in the first of Eq.~(\ref{newtongrav}), and use Eq.~(\ref{divH}), we obtain
\[
-4\pi G \rho_\mathrm{B} \!\stackrel{(\ref{newtongrav})}{=}\!
\boldsymbol\nabla\bullet \boldsymbol g_\text{N} \!\stackrel{(\ref{helmholtz})}{=}\!
\boldsymbol\nabla\bullet \boldsymbol F - \boldsymbol\nabla\bullet \boldsymbol H \!\stackrel{(\ref{divH})}{=}\!
\boldsymbol\nabla\bullet \boldsymbol F \!\stackrel{(\ref{interpolation})}{=}\!
\boldsymbol\nabla\bullet \mu\bigg(\frac{g_\text{M}}{a_0}\bigg) \boldsymbol g_\text{M}
.
\]
Equation (\ref{rotg}) states that the field is curl-free. Hence, $\boldsymbol g_\text{M}$ is a conservative field with a potential. When we substitute $\boldsymbol g_\text{M}=-\boldsymbol\nabla\phi$, we find Eq.~(\ref{PDE}). This proves that the system Eqs.\ (\ref{newtongrav})-(\ref{interpolation}) implies Eq.~(\ref{PDE}). For the converse, we note that when we have a solution of Eq.~(\ref{PDE}), we can calculate $\boldsymbol g_\text{M}$. This automatically satisfies Eq.~(\ref{rotg}). We can then define $\boldsymbol F$ with Eq.~(\ref{interpolation}). We then let Eq.~(\ref{helmholtz}) be its Helmholtz decomposition. This defines the functions $\boldsymbol g_\text{N}$ and $\boldsymbol H$, which must satisfy the second of Eq.~(\ref{newtongrav}) and Eq.~(\ref{divH}), stating that the first component is curl-free and the second component is divergence-free. This completes the proof.

In any numerical code that uses this system, the linear PDEs can be solved using standard numerical methods for linear equations. The nonlinear equation requires a special solver. We have devised the following numerical iteration scheme that approaches the solution of the system (\ref{newtongrav})-(\ref{interpolation}). We first solve Eq.~(\ref{newtongrav}) for the Newtonian gravity field $\boldsymbol g_\text{N}$, which we store in memory. Then we take $\boldsymbol H=\boldsymbol 0$ for the initial value. The iteration scheme consists of the following steps:
\begin{enumerate}
\item
Calculate $\boldsymbol F$ with 
Eq.~(\ref{helmholtz}).
\item
Calculate $\boldsymbol g_\text{M}$ with 
Eq.~(\ref{inverse}).
\item
Project $\boldsymbol g_\text{M}\longrightarrow \boldsymbol g_{\text{M}\parallel}$ onto the curl-free part, as per (\ref{rotg}).
\item
Calculate $\boldsymbol F$ with Eq.~(\ref{interpolation}).
\item
Calculate $\boldsymbol H$ with 
Eq.~(\ref{helmholtz}).
\item
Project $\boldsymbol H\longrightarrow\boldsymbol H_\perp$ onto the divergence-free part, as per (\ref{divH}).
\item
Repeat: Go to step 1.
\end{enumerate}
The calculation in steps 3 and 6 can quickly be made in the Fourier domain (see Appendix \ref{AppB} for the equations). The use of the fast Fourier transform (FFT) allows us to perform these steps fast.

The iterative procedure converges to a numerically stable solution. Because we start with a zero $\boldsymbol H$ field, its change after the iterations allows us to estimate the error in the numerical solution. It was found that the error initially drops with a factor of $10$ per iteration, but after about 15 iterations, 5 iterations are required for the error to drop with a factor $10$, until machine precision is reached \citep[see][where it was used for the first time]{Platschorre2019}. One clue to the explanation for the fast convergence is the following. For a spherical field, we have $\boldsymbol H=\boldsymbol 0$ \citep[see][]{Bekenstein1984}. Because the masses are spherical, close to a mass, the solution is approximately spherically symmetric. Far away from a collection of masses, the solution becomes spherical as well. Thus, near the masses and away from the masses, the initial value is correct. In the iterative steps, the field in the intermediate regions is changed. Here, $\boldsymbol H$ is not zero.

The apparent matter (baryonic plus dark matter) distribution can be calculated by substituting the correct potential (the potential that is consistent with the observed particle motions and lensing) in the incorrect Newton-Poisson equation. This is equivalent to
\begin{equation}
\rho_\mathrm{B}(\boldsymbol r) = \frac{-1}{4\pi G} \boldsymbol\nabla \bullet \boldsymbol g_\text{N}
, \quad
\rho_\mathrm{A}(\boldsymbol r) = \frac{-1}{4\pi G} \boldsymbol\nabla \bullet \boldsymbol g_\text{M}
.
\label{apparentdensity}
\end{equation}
The apparent density $\rho_\text{A}$ is also known as the effective density \citep[see][]{Banik2022}. This leads to an apparent dark matter density of $\rho_\text{A}-\rho_\text{B}$, also called phantom dark matter density. The total baryonic mass and apparent mass in a volume $W$ can be found from the flux through its boundary,
\begin{equation}
M_\text{B}(W) = -{\oiintop}_{\partial W} \boldsymbol g_\text{N} \bullet \mathrm d\boldsymbol S
, \quad
M_\text{A}(W) = -{\oiintop}_{\partial W} \boldsymbol g_\text{M} \bullet \mathrm d\boldsymbol S
.
\label{totalmass}
\end{equation}

\subsection{An equivalent system for QuMOND}
An alternative to the AQUAL formalism of MOND is quasi-linear MOND or  QuMOND \citep{Milgrom2010,Candlish2014,Lughausen2015,Nagesh2022}. This model can be cast as a similar system
\begin{empheq}[left=\empheqlbrace]{align}
\boldsymbol\nabla \bullet \boldsymbol g_\text{N} &= -4\pi G \rho_\mathrm{B}
, &
\boldsymbol\nabla \times \boldsymbol g_\text{N} &= \boldsymbol 0
\nonumber
, & \\
\boldsymbol g_\text{M} + \boldsymbol H &= \nu\bigg(\frac{g_\text{N}}{a_0}\bigg) \boldsymbol g_\text{N}
, &
\boldsymbol\nabla \times \boldsymbol g_\text{M} &= \boldsymbol 0 , & \boldsymbol\nabla \bullet \boldsymbol H = 0
\nonumber
.
\label{interpolation}
\end{empheq}
We do not consider QuMOND any further. However, the algorithm can easily be adapted using the same steps.

\section{The algorithm: Dealing with the self-gravity}
\label{sect3}
We now describe the implementation of the method. The code needs to numerically approximate the solution of the MOND PDEs in Eqs.\ (\ref{newtongrav})-(\ref{interpolation}) and propagate particle motion and a time-dependent mass density. The particle equations of motion Eq.~(\ref{ddr}) are implemented with Leapfrog,
\begin{equation}
\boldsymbol v_i \longrightarrow \boldsymbol v_i + \boldsymbol g_i(t)\Delta t , \quad
\boldsymbol r_i \longrightarrow \boldsymbol r_i + \boldsymbol v_i\Delta t , \quad
i=1\ldots N
.
\label{motion}
\end{equation}
Although the formulas do not explicitly show this, the velocities and positions represent the values at a half time step difference. In principle, an actual half time step needs to be made initially (and also finally) to obtain the true Leapfrog integration scheme.

The code uses a discrete Fourier transform, so that space is represented as a periodic cubic lattice of $n^3$ cells we call pixels. Hence, the scheme does not have nested grids; the grid cells are all equal and do not change. However, due to the periodic nature of the simulated space, matter inside one cubic volume is effectively influenced by the gravity of the neighboring copies. This effect is nonphysical, and a volume of empty space needs to surround the matter when this effect is to be small. Ideally, the center of mass is placed in the origin, and it is ensured that particles stay away from the boundary of the cube.

\subsection{Interpolation for the particle density function}
The main drawback of the particle-mesh (PM) method is the lack of accuracy at the short-distance scale of the pixels. Although each particle is located in a single pixel, we desire accuracy that is below the grid size. This influences both the field created by and the acceleration on the particle. First, the field created by particle $i$ close to the mass should resemble a spherical field about the point $\boldsymbol r_i$, which is not the center of a pixel. Second, we also need to approximate the acceleration on particle $i$ by the field at $\boldsymbol r_i$, for which the value at the center of the pixel is not accurate.

\begin{figure}[ht]
\centering
\includegraphics{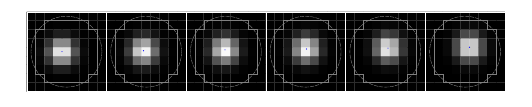}
\caption{Anti-aliasing needed to suppress the self-gravity. A cross section of the Gaussian density Eq.~(\ref{smooth}) with one-pixel variance about (blue) points in the middle grid cell, representing a 12th order weight function. We only account for the pixels inside a sphere with a radius of $4s$ (dashed circles). This results in a domain of 251 cells (inside the contours). The linear motion of a point particle across the middle cell (left to top right) shows that the simulated density changes smoothly between panels.}
\label{FigureFig1}
\end{figure}

We used a Gaussian kernel to create a smoothed density in order to suppresses anti-aliasing. The density of Eq.~(\ref{density}) was replaced by
\begin{equation}
\rho_\text{B}(\boldsymbol r) =
\frac{1}{s^3(2\pi)^{3/2}} \sum_{i=1}^N m_i\ \text{exp}\Bigg(\frac{-|\boldsymbol r-\boldsymbol r_i|^2}{2s^2}\Bigg)
.
\label{smooth}
\end{equation}
Hence, each of the $N$ point particles is represented by the Gaussian in Eq.~(\ref{smooth}) centered about the particle position. We chose a standard deviation with a width of one pixel $s$. The Gaussian is then effectively a cloud-in-cell (CIC) method of order $12$ \citep[see][]{Hockney1988}. This is explained in Appendix \ref{AppA}. Because Gaussians drop off fast, we implemented the calculation of the density by only addressing the pixels that are within a distance of $4s$ from the central pixel containing the particle. To do this, we stored a list with the $251$ coordinates of the ball-shaped domain with a radius of $4s$. This is illustrated in Fig.~\ref{FigureFig1}. Moreover, we created a lookup table for the exponential function.

\subsection{Interpolation for the forces on the particles}
Next, we describe how we approximated the acceleration of particle $i$. Again, we wished to know the acceleration at a point $\boldsymbol r_i$ inside some pixel with subpixel accuracy. However, the acceleration $\boldsymbol g_\text{B}$ is only calculated at the pixel positions. We would therefore need to interpolate between the values at different pixels. For the force on particle $i$, located at $\boldsymbol r_i$, we used the following smoothed average:
\begin{equation}
\boldsymbol g_i =
\frac{-1}{s^3(2\pi)^{3/2}} \iiintop \boldsymbol \nabla \phi(\boldsymbol r)
\text{exp}\bigg( \frac{-|\boldsymbol r-\boldsymbol r_i|^2}{2s^2} \bigg) \mathrm dx\ \mathrm dy\ \mathrm dz
.
\label{gi}
\end{equation}
On the mesh, the gradient of the potential $\boldsymbol\nabla\phi$ in this expression is calculated with the finite-difference gradient, and the integral in this expression becomes a pixel sum. We only summed over pixels within a distance $4s$ from the pixel containing $\boldsymbol r_i$.

The smoothing methods discussed in this section will generally improve the accuracy in any PM method. However, in the case of MOND, this smoothing is essential for dealing with the self-gravity. The software is available on GitHub\footnote{\texttt{MONDPMesh} codebase: \url{https://github.com/Joost987/MONDPMesh}.} under the MIT license and is archived in the ASCL \citep{DeNijs2023github}.

\section{Results: Validation using analytic solutions}
\label{sect4}
The problem of self-gravity dominating the actual force on a point particle is inherent to the PM method for MOND. The effect of an imprecise subpixel resolution for the position of a particle and the consequently imprecise evaluation of the acceleration on that particle is always present, even for a single particle. We therefore also need to test the accuracy of the method for linear motion. We have found that in our simulations, a point particle of mass $m$ experiences nonphysical accelerations with the average value of $1.08\cdot 10^{-3}\sqrt{Gma_0}/s$ for a $256^3$ grid \citep[see][]{DeNijs2023}. This small error is independent of the number of pixels.

In this section, we present results of simulations with the code for three systems: the two-body problem, a ringed system, and an isotropic three-dimensional system. Because the main interest is in MOND, we considered the deep MOND case. This means that the interpolation functions we used in our simulation are
\[
\mu(x) = x , \quad \nu(y) = \frac{1}{\sqrt{y}}
.
\]
For this regime, \citet{Milgrom1994,Milgrom1997,Milgrom2014} found, using the method of the virial, that the kinetic energy for a bound system with the center-of-mass motion at rest is
\begin{equation}
E_\text{kin} =
\frac{M\overline{\boldsymbol v\bullet\boldsymbol v^{\vphantom{2}}}}{2} =
\frac{M\overline{v^2}}{2} = \frac{\sqrt{Ga_0}}{3} \bigg( M^{3/2} - \sum_i {m_i}^{3/2} \bigg)
,
\label{virial}
\end{equation}
with $m_i$ the individual particle masses, and $M$ the total mass. If there is no external field, this single equation allows us to calculate the MOND forces between particles for highly symmetric cases, thereby bypassing the nonlinear PDE. We used these analytic solutions to test our PM code.

\subsection{Example I: Wide binary system}
\label{sect4.1}
We considered the two-body problem in MOND. This is a model for a binary stellar system with two comparable masses with a wide separation. The interparticle force is found from Eq.~(\ref{virial}) to be
\begin{equation}
F_{12} = \frac{2\sqrt{Ga_0}}{3r} \Big( M^{3/2} - m_1^{3/2} - m_2^{3/2} \Big)
,
\label{twobodyforce}
\end{equation}
with $M=m_1+m_2$, and where $r=|\boldsymbol r_2-\boldsymbol r_1|$ is the separation. We can rewrite the mass dependence using
\begin{align}
M^{3/2} - m_1^{3/2} - m_2^{3/2} &= \textstyle m_1\Big( \sqrt{M_{\phantom{0}}}-\sqrt{\vphantom{M}m_1} \Big) + m_2\Big(\sqrt{M_{\phantom{0}}}-\sqrt{\vphantom{M}m_2}\Big)
\nonumber \\
&= m_1m_2\bigg( \frac{1}{\sqrt{M_{\phantom{0}}}+\sqrt{\vphantom{M}m_1}} + \frac{1}{\sqrt{M_{\phantom{0}}}+\sqrt{\vphantom{M}m_2}} \bigg)
.
\label{rewrite}
\end{align}
The equations of motion of the two particle system are
\begin{empheq}[left=\empheqlbrace]{align}
\dot{\boldsymbol r}_1 &= \boldsymbol v_1 , \nonumber \\
\dot{\boldsymbol r}_2 &= \boldsymbol v_2 , \nonumber \\
\dot{\boldsymbol v}_1 &= \frac{2\sqrt{Ga_0}}{3} \frac{M^{3/2} - m_1^{3/2} - m_2^{3/2}}{m_1} \frac{\boldsymbol r_2-\boldsymbol r_1}{|\boldsymbol r_1-\boldsymbol r_2|^2} , \nonumber \\
\dot{\boldsymbol v}_2 &= \frac{2\sqrt{Ga_0}}{3} \frac{M^{3/2} - m_1^{3/2} - m_2^{3/2}}{m_2} \frac{\boldsymbol r_1-\boldsymbol r_2}{|\boldsymbol r_1-\boldsymbol r_2|^2}
.
\label{system}
\end{empheq}
\begin{figure}[ht]
\centering
\includegraphics{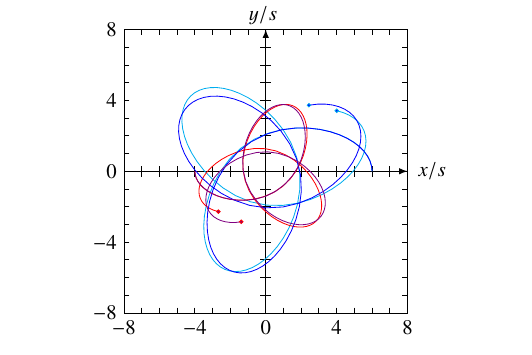}
\caption{Simulated orbits of the two-body system projected onto the $xy$ plane for the mass ratio $m_1/m_2=3/2$. The mesh has a size of $128^3$, and there are $100$ time steps, each with four iterations for the MOND force. Red ($m_1$) and cyan ($m_2$) are the orbits obtained from numerical integration of Eqs.(\ref{system}). Purple and blue are the orbits calculated with the PM code. 
The difference between the PM code and the ODEs is a slow dephasing of the orbits.}
\label{Figure2}
\end{figure}
A numerical solution of this system of ordinary differential equations (ODE) is shown in Fig.~\ref{Figure2}. Here, it is compared with the results of our PM code, showing that the two methods agree reasonably well, but diverge ultimately. The system has the following constants of motion:
\begin{align*}
& \boldsymbol P = m_1\boldsymbol v_1 + m_2\boldsymbol v_2 , \\
& \boldsymbol R(t) - \frac{\boldsymbol P}{M}t = \frac{m_1\boldsymbol r_1 + m_2\boldsymbol r_2}{M} - \frac{\boldsymbol P}{M}t , \\
& \boldsymbol R(t) \times \boldsymbol P , \\
& \boldsymbol L = m_1\boldsymbol r_1\times \boldsymbol v_1 + m_2\boldsymbol r_2\times \boldsymbol v_2 , \\
& E = \frac{m_1v_1^2+m_2v_2^2}{2} + 2\sqrt{Ga_0} \frac{M^{3/2}-m_1^{3/2}-m_2^{3/2}}{3} \log|\boldsymbol r_1-\boldsymbol r_2|
.
\end{align*}
Because the PM is cubic, which breaks the local spherical symmetry, we expect that the total angular momentum does not remain constant in the simulations. How well angular momentum is conserved by our PM code is shown in Fig.~\ref{Figure3}. If the center of mass is located at the origin, $\boldsymbol R=\boldsymbol 0$, the single-particle angular momenta $\boldsymbol L_1$ and $\boldsymbol L_2$ are also constant. However, the numerical error in $\boldsymbol L_1$ and $\boldsymbol L_2$ becomes large when the particle is within a distance of one pixel from the origin. The reason is that the center of mass may have drifted away from the origin by one pixel. When we subtract $\boldsymbol R$ from $\boldsymbol r_1$ and $\boldsymbol r_2$, the errors disappear. For grid sizes $128^3$, $256^3$, and $512^3$, the error in total angular momentum is $4\%$, $2\%$, and $1\%$, respectively.

\begin{figure}[ht]
\centering
\includegraphics{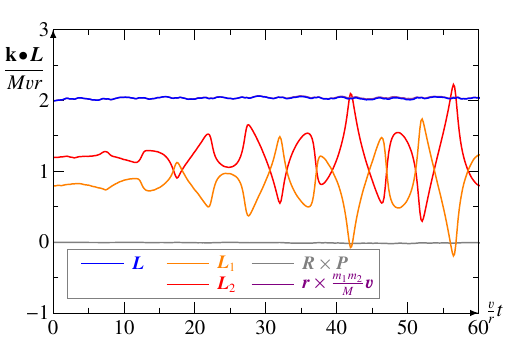}
\caption{Angular momentum vs.\ time $t$ for the simulation shown in Fig.~\ref{Figure2}. These are constants of motion of the ODE system Eq.~(\ref{system}) and numerically constant (with machine precision, not shown) when integrated with Leapfrog. In the PM code, the total angular momentum (blue) is well conserved, but the single-particle angular momenta (orange and red) have large errors when the distance to the origin is smaller than one pixel.}
\label{Figure3}
\end{figure}
\begin{figure}[ht]
\centering
\includegraphics{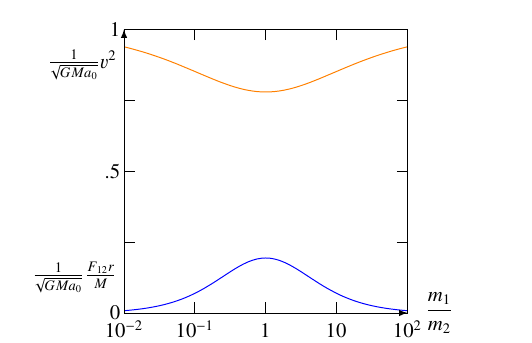}
\caption{Force times radius (blue) and squared velocity (orange) vs.\ mass ratio for the two-body system. The horizontal axis is logarithmic, making the curves symmetric about $m_1/m_2=1$.}
\label{Figure4}
\end{figure}

\begin{figure}[ht]
\centering
\includegraphics{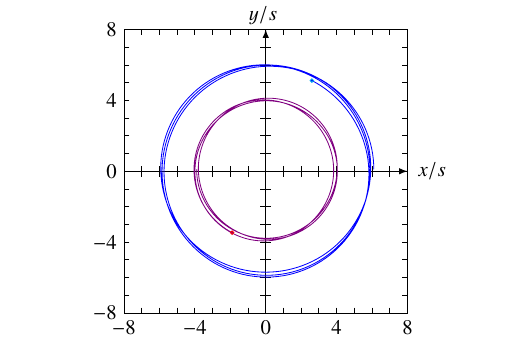}
\caption{Simulated orbits of two bodies with a mass ratio $m_1/m_2=3/2$ as in Fig.~\ref{Figure2}, but with 400 time steps. The initial velocities are given by Eq.~(\ref{twobodyvelocity}) and should result in circular orbits. These are plotted in Fig~\ref{Figure4}. This simulation thus confirms the force formula Eq.~(\ref{twobodyforce}) for the deep MOND case and validates the algorithm. It also illustrates how well the algorithm performs on the subpixel scale.}
\label{Figure5}
\end{figure}

The solution with the two particles in circular orbit about the center of mass is
\begin{align*}
\boldsymbol r_1 &= \frac{m_2r}{M}(\mathbf i\cos\omega t+\mathbf j\sin\omega t) , &
\boldsymbol r_2 &= \frac{-m_1r}{M}(\mathbf i\cos\omega t+\mathbf j\sin\omega t) , \\
\boldsymbol v_1 &= \frac{-m_2v}{M}(\mathbf i\sin\omega t-\mathbf j\cos\omega t) , &
\boldsymbol v_2 &= \frac{m_1v}{M}(\mathbf i\sin\omega t-\mathbf j\cos\omega t)
,
\end{align*}
where $r=|\boldsymbol r_2-\boldsymbol r_1|=r_1+r_2$ is the separation, $v=|\boldsymbol v_2-\boldsymbol v_1|$ is the relative velocity, and $\omega$ is the orbital frequency. With Huygens' law of the centripetal force applied to the motion of the reduced coordinate $r$, we obtain for the force
\[
\frac{m_1v_1^2}{r_1} = \frac{m_2v_2^2}{r_2} = \frac{m_1m_2v^2}{Mr} = F_{12}
,
\]
and for the kinetic energy
\[
\frac{m_1v_1^2+m_2v_2^2}{2} = \frac{m_1m_2v^2}{2M} =
\frac{M\overline{v^2}}{2} = E_\text{kin}
.
\]
Using Eqs.~(\ref{virial})-(\ref{rewrite}), we find that the velocity $v$ of the reduced motion and the orbital frequency $\omega$ are given by the expression
\begin{equation}
v^2 = (\omega r)^2 = \frac{M^2\overline{v^2}}{m_1m_2} =
\frac{2\sqrt{GMa_0}}{3}\bigg( \frac{1}{1+\sqrt{m_1/M}} + \frac{1}{1+\sqrt{m_2/M}} \bigg)
.
\label{twobodyvelocity}
\end{equation}
The velocities are independent of the separation. The theoretical dependence of the velocity on the mass ratio is plotted in Fig.~\ref{Figure4}. Figure \ref{Figure5} shows the numerical simulation of our PM code for two particles with the initial relative velocity from Eq.~(\ref{twobodyvelocity}). It convincingly shows that motion is circular, which verifies Milgrom's two-particle force law. It also illustrates that we can obtain reasonable subpixel resolution.

\subsection{Example II: Ring galaxy}
\begin{figure}[tb]
\centering
\includegraphics{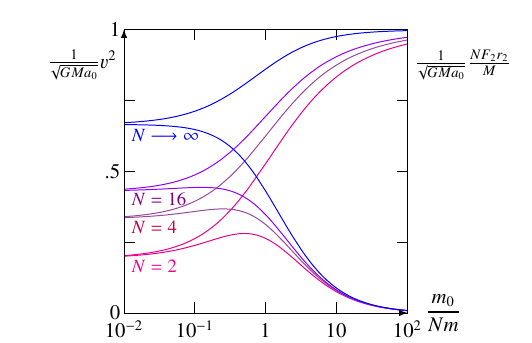}
\caption{Force times radius (lower curves) and squared velocity (higher curves) vs.\ the mass ratio for particles on a regular $N$-gon orbiting a central body in circular motion. From Eqs.~(\ref{ringforce})-(\ref{ringvelocity}), where $m_0$ is the central mass and $Nm$ is the ring mass.}
\label{Figure6}
\end{figure}

\begin{figure}
\centering
\includegraphics{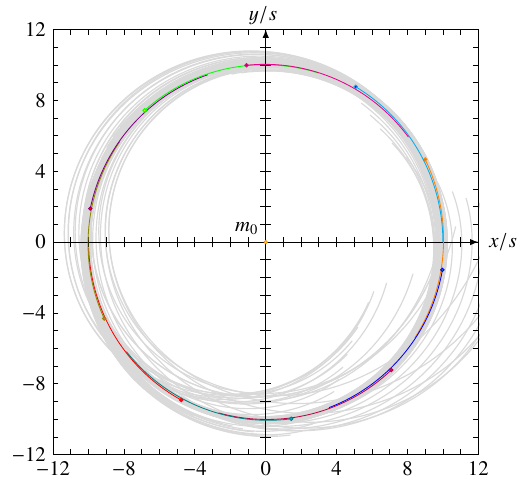}
\caption{Simulated orbits for a ring of $N=100$ particles with a mass $m$ around a central mass $m_0$ for a mass ratio of $m_0/Nm=3/2$ as in Fig.~\ref{Figure2}. The orbits of the first 80 time steps of 10 particles are shown in color. In gray we plot 576 steps for all particles. The initial velocities are given by Eq.~(\ref{ringvelocity}), plotted in Fig~\ref{Figure6}, which in theory results in circular orbits. Although this confirms the force formula Eq.~(\ref{ringforce}), numerical round-off errors on the subpixel scale propagate and cause the orbits to diverge.}
\label{Figure7}
\end{figure}

Again, we tested the code with a system that was solved analytically by \citet{Milgrom1994}: a central mass $m_0$, surrounded by a ring with a radius $r_2$ with $N$ particles with a mass $m$ (i.e., the system has $N+1$ particles). A beautiful realization of such a ring galaxy is Hoag's Object \citep{Schweizer1987}. The central mass is at rest at the origin. We created a rotating ring of $N$ particles on a regular $N$-gon ($N\geq 2$),
\begin{align*}
\boldsymbol r_i &= r_2 \mathbf i\cos(\omega t+2\pi i/N) + r_2 \mathbf j\sin(\omega t+2\pi i/N)
, \\
\boldsymbol v_i &= -v\mathbf i\sin(\omega t+2\pi i/N) + v\mathbf j\cos(\omega t+2\pi i/N)
.
\end{align*}
The force on the particles in the ring is given by
\begin{equation}
F_2 = \frac{2\sqrt{Ga_0}}{3Nr_2} \Big( M^{3/2} - m_0^{3/2} - Nm^{3/2} \Big)
.
\label{ringforce}
\end{equation}
Here, $M=m_0+Nm$. We can now write
\begin{align*}
M^{3/2} - m_0^{3/2} &= \textstyle m_0\Big( \sqrt{M_{\vphantom{0}}}-\sqrt{\vphantom{M}m_0}\Big) + Nm\sqrt{M_{\vphantom{0}}} \\
&= MNm \Bigg( \frac{1}{\sqrt{M_{\vphantom{0}}}+\sqrt{\vphantom{M}m_0}} + \frac{\sqrt{m_0}}{M} \Bigg)
.
\end{align*}
The velocity of the ring particles and the orbital frequency are found by equating the force Eq.~(\ref{ringforce}) to $F_2=mv^2/r_2=m\omega^2r_2$,
\begin{equation}
v^2 = 
\frac{M\overline{v^2}}{Nm} = \frac{2\sqrt{GMa_0}}{3} \Bigg( \frac{1}{1+\sqrt{m_0/M}} + \sqrt{m_0/M} - \sqrt{m/M} \Bigg)
.
\label{ringvelocity}
\end{equation}
The force times radius and squared velocity is plotted in Fig.~\ref{Figure6}. The result of the simulation for $N=100$ particles is plotted in Fig.~\ref{Figure7}. If the mass of the ring is not very low compared to the central mass, the system is unstable, and the ring breaks apart after a few orbits.


\subsection{Eample III: The isothermal sphere in MOND}
\label{sect3.3}
Our third application is the simulation of a spherically symmetric system in thermodynamic equilibrium. This could be a globular cluster with up to $10^6$ stars or a galaxy cluster with up to $10^3$ galaxies. These system are in the deep MOND regime or in the crossover regime (accelerations around $a_0$).

A system of $N$ point particles with equal mass $m$ in hydrostatic equilibrium with a temperature $T$ has a velocity dispersion given by
\begin{equation}
\frac{3k_\mathrm{B}T}{2m} =
\frac{\overline{\boldsymbol v\bullet\boldsymbol v^{\vphantom{2}}}}{2} = \frac{\overline{v^2}}{2} = \frac{\sqrt{GMa_0}}{3}
.
\label{vsqsuare}
\end{equation}
The last equality is the thermodynamic limit of Eq.~(\ref{virial}). For a spherical system, the particle number density in terms of the potential is given by
\begin{equation}
\frac{n(r)}{n(0)} =
\exp \bigg( \frac{-m\phi(r)}{k_\mathrm{B}T} \bigg) =
\exp \bigg( \frac{-3\phi(r)}{\overline{v^2}} \bigg) =
\exp \bigg( \frac{-9\phi(r)}{2\sqrt{GMa_0}} \bigg)
.
\label{numberdensity}
\end{equation}
Here, we used hydrostatic equilibrium and the ideal gas law. Substitution of the mass density $\rho_\text{B}(r)=mn(r)$ and Eq.~(\ref{numberdensity}) in the MOND Eq.\ 
(\ref{PDE}) gives the following equation for the potential:
\begin{equation}
\frac{1}{a_0r^2} \frac{\mathrm d}{\mathrm dr} \bigg(r\frac{\mathrm d\phi}{\mathrm dr} \bigg)^2 =
4\pi G\rho_\text{B}(0)\exp \bigg( \frac{-9\phi(r)}{2\sqrt{GMa_0}} \bigg)
.
\label{isothermal}
\end{equation}
\begin{figure}
\centering
\includegraphics{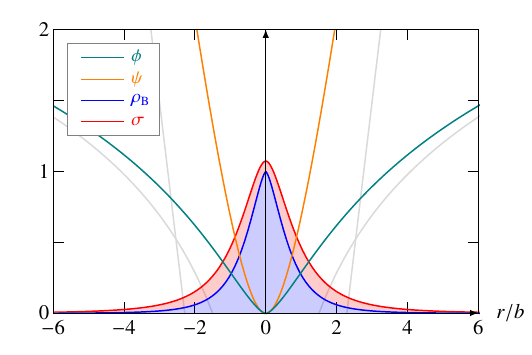}
\caption{Potentials and mass distributions for the isothermal state. In teal and orange we show the potential $\phi(r)$ as given by Eq.~(\ref{phi}) and the deflection potential $\psi=\int \phi\mathrm dz$ for the line of sight passing the center at distance $r$, determining weak gravitational lensing. The asymptotes are shown (in gray) and the units are $\sqrt{GMa_0}b^j$, $j=0,1$. In blue and red we show the density $\rho_\text{B}(r)$ as given by Eq.~(\ref{rho}) and the surface mass density $\sigma=\int \rho_\text{B}\mathrm dz$ in units of $3M/4\pi b^j$, $j=3,2$.}
\label{Figure8}
\end{figure}
The solution of this equation is found to be
\begin{align}
\phi(r) &= \frac{2\!\sqrt{GMa_0}}{3}  \log\bigg( 1 + \frac{r^{3/2}}{b^{3/2}} \bigg)
\label{phi}
, \\
\boldsymbol g_\text{M}(\boldsymbol r) &= -\boldsymbol r \sqrt{\dfrac{GMa_0}{b^3r}}\ \bigg( 1 + \dfrac{r^{3/2}}{b^{3/2}} \bigg)^{-1}
\label{g}
, \\
M(r) &= \frac{Mr^3}{b^3} \bigg(\displaystyle 1 + \frac{r^{3/2}}{b^{3/2}} \bigg)^{-2}
\label{M}
, \\
\rho_\text{B}(r) &= 
\frac{3M}{4\pi b^3}\ \bigg(\displaystyle 1 + \frac{r^{3/2}}{b^{3/2}} \bigg)^{-3}
\label{rho}
.
\end{align}
Here, $b$ is a parameter with the dimension of a length. When the function in Eq.~(\ref{phi}) is substituted in Eq.~(\ref{isothermal}), we can solve for this parameter $b$ in terms of the mass $M$ and the central density $\rho_\text{B}(0)$. Hence, we have found a two-parameter family of solutions. The function $M(r)$ in Eq.~(\ref{M}) represents the mass inside a sphere of radius $r$, and $M=\lim_{r\rightarrow\infty}M(r)$ is the total mass. The physical meaning of the parameter $b$ becomes clear when we calculate the mass of a homogeneous sphere with radius $b$, with the constant density equal to the value $\rho(0)$ at the center of the isothermal distribution, given in Eq.~(\ref{rho}). This mass is precisely equal to the total mass
\[
M=
4\pi \int\limits_0^\infty \rho_\text{B}(r)r^2\mathrm dr =
4\pi \int\limits_0^b \rho_\text{B}(0)r^2\mathrm dr = \frac{4\pi b^3}{3}\rho_\text{B}(0)
.
\]
The density is finite at the origin and also drops off sufficiently fast to zero to allow normalization: No cutoff is needed to keep the total mass finite. These physically convenient properties are absent from the isothermal solution in Newtonian gravity. The potential and density and the resulting deflection potential and surface-mass density are plotted in Fig.~\ref{Figure8}. Kinetic energy and potential energy are, with Eq.~(\ref{vsqsuare}),
\[
E_\mathrm{kin} = \frac{M\overline{v^2}}{2} , \quad
E_\mathrm{pot} = 4\pi \int\limits_0^\infty \rho_\text{B}(r)\phi(r) r^2\mathrm dr = \frac{3M\overline{v^2}}{2} 
.
\]
The solution given in Eqs.~(\ref{phi})-(\ref{rho}) is applicable for $g_\text{M}\ll a_0$. Because the maximum acceleration is found at $r=b/2^{2/3}$, the requirement becomes $GM/b^2\ll 9a_0/2^{4/3}$. The Virgo cluster, for example, satisfies this. Equation (\ref{rho}) could be a realistic description for spherical gas clouds, elliptical galaxies, and clusters.

We derived the formula for the apparent mass density, which would give the apparent dark matter distribution function. Equations (\ref{apparentdensity})-(\ref{totalmass}) give
\begin{align*}
M_\text{A}(r) &=
\sqrt{\dfrac{Ma_0}{G}} r\ \bigg( 1 + \dfrac{b^{3/2}}{r^{3/2}} \bigg)^{-1}
, \\
\rho_\text{A}(r) &=
\frac{1}{4\pi} \sqrt{\dfrac{Ma_0}{Gb^3r}}\ \bigg( \frac{5}{2} + \dfrac{r^{3/2}}{b^{3/2}} \bigg) \bigg( 1 + \dfrac{r^{3/2}}{b^{3/2}} \bigg)^{-2}
.
\end{align*}
The total apparent mass diverges as $r\longrightarrow\infty$. Whereas the actual density $\rho_\text{B}$ is smooth at the center, the apparent mass density $\rho_\text{A}$ has a cusp at the core. The isothermal solution shows typical deep MOND behavior: Only the total baryonic mass $M$ determines the velocity distribution, with a central density that can be varied independently. The lower the density at the center, the larger the system system size and the higher the apparent dark matter mass.

\begin{figure}[tb]
\vspace{-2mm}
\centering
\includegraphics{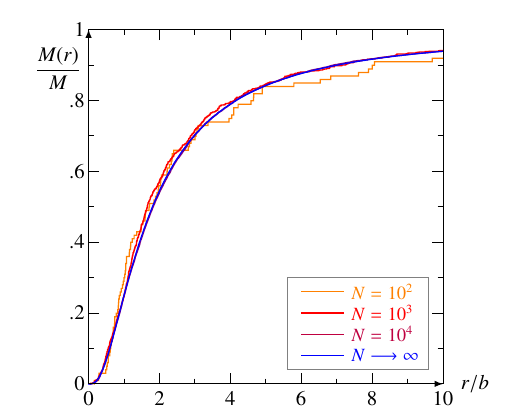}
\caption{Cumulative mass $M(r)$ of the spherical isothermal galaxy vs.\ $r$. The (free) parameter $b$ determines the galaxy size. The (blue) curve is the theoretical result given by Eq.~(\ref{M}). The realizations (orange, red, and purple) are calculated from Eq.~(\ref{boxmuller}).}
\label{Figure9}
\end{figure}

\begin{figure}[tb]
\vspace{-2mm}
\centering
\includegraphics{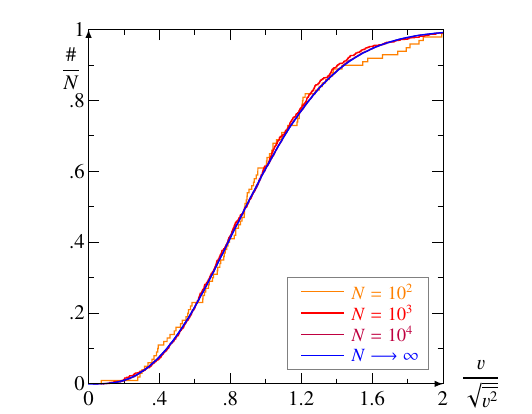}
\caption{Cumulative velocity distribution for the spherical isothermal galaxy. This is the fraction $\#(\leq\! v)/N$ of particles with velocities below $v$ vs.\ the value $v$ of this upper bound. The realizations (orange, red, and purple) are calculated from Eq.~(\ref{boxmuller}).}
\label{Figure10}
\end{figure}

\begin{figure*}[ht]
\centering
\includegraphics{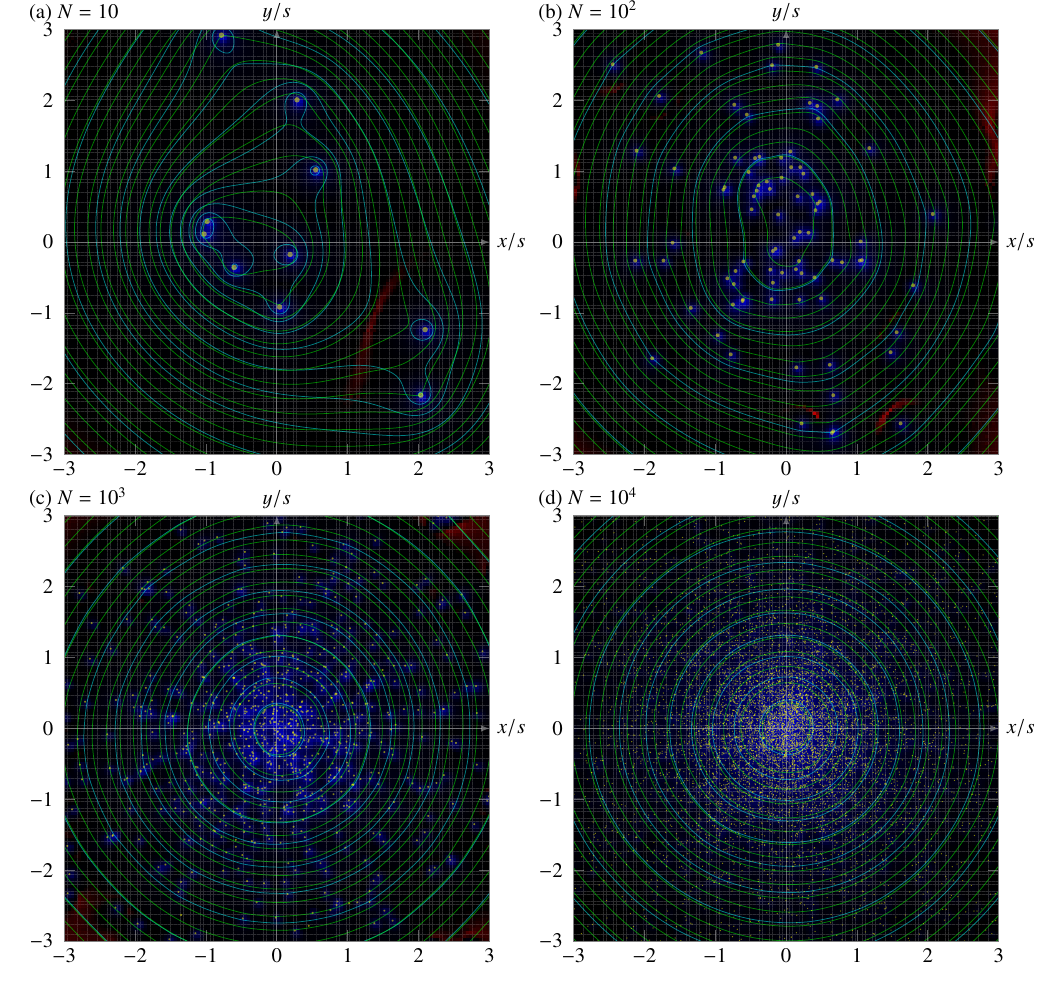}
\caption{Surface-mass densities $\sigma(x,y)$ and deflection potentials $\psi(x,y)$ for a globular cluster from the central $128\times 128$ part of a simulated cube of $256^3$ pixels. The yellow dots are the stars that define the actual density. The resulting apparent dark matter surface density is plotted in blue according to Eq.~(\ref{apparentdensity}). Negative value are shown in red \citep{Milgrom1986negativemass}. They are found for $N\leq 10^2$. The equipotential curves of the deflection potential for Newtonian and Milgromian gravity are plotted in green and cyan. Although the densities are far from circularly symmetric, the deflection potential is nearly circular for $N\geq 10^3$. This indicates that the retrieval of a clumpy mass distribution from weak lensing may be problematic for large $N$.}
\label{Figure11}
\end{figure*}

We now study the dynamics of the isothermal solution with our PM code. We need the cumulative distribution function for the radial probability density. According to Eq.~(\ref{M}), this is
\[
\frac{M(r)}{M} =
\bigg( \displaystyle 1 + \dfrac{b^{3/2}}{r^{3/2}} \bigg)^{-2}
.
\]
In order to obtain random realizations, our point masses were found from seven random numbers $\xi_1\in (-1,1)$, $\eta_1,\eta_2,\eta_3\in (0,1)$, $\zeta_1,\zeta_2,\zeta_3\in (0,2\pi)$, sampled homogeneously from these intervals and the formulae
\begin{align}
\boldsymbol r_i &= b \bigg( \sqrt{1-\xi_1^2} \Big(\mathbf i\cos\zeta_1 + \mathbf j \sin\zeta_1 \Big) + \mathbf k\xi_1 \bigg) \bigg( \frac{1}{\sqrt{\eta_1}}-1 \bigg)^{-2/3}
,
\label{boxmuller} \\
\boldsymbol v_i &= \sqrt{-\tfrac{2}{3}\overline{v^2} \log\eta_2} \Big( \mathbf i\cos\zeta_2 + \mathbf j \sin\zeta_2 \Big) + \sqrt{-\tfrac{2}{3}\overline{v^2} \log\eta_3} \mathbf k\cos\zeta_3
.
\nonumber
\end{align}
The Box-Muller transform was used to generate the Gaussian values of the Maxwell-Boltzmann velocity distribution. Figures \ref{Figure9} and \ref{Figure10} show how close the realizations are to the cumulative distributions for the radial coordinate and for the velocity. Figure \ref{Figure11} shows the realizations projected onto the celestial plane and the resulting apparent dark matter surface density. For $N=10^3$ or larger, the potential is nearly spherical, and it is well described with our analytic solution.

\begin{figure}[t]
\centering
\includegraphics{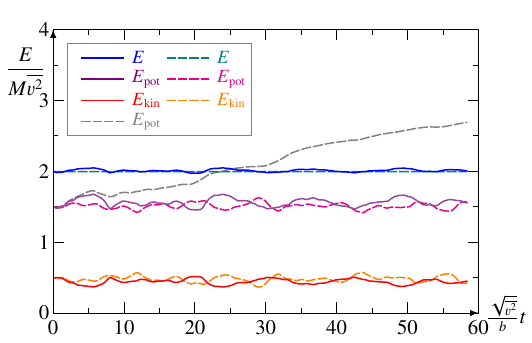}
\caption{Energies vs.\ time for $N=100$ particles on a $256^3$ grid. The kinetic, potential, and total energy for the self-consistent interacting system of the particles from the PM code are shown as solid curves (red, purple, and blue). The kinetic, potential, and total energy from the ODE system of the particles in the external potential are shown as dashed curves (orange, violet, and teal). The fluctuations in the total energies are purely numerical, and the other energies have thermal fluctuations. The dashed gray curve is $E_\text{pot}=\int \rho_\text{B}\phi\mathrm dx\mathrm dy\mathrm dz$ with the PM matter density and Eq.~(\ref{phi}), hence substituting a finite $N$ solution in the potential obtained in the thermodynamic limit. We subtracted the self-energies.}
\label{Figure12}
\end{figure}

We solved the evolution of $N$ particles in the external potential Eq.~(\ref{phi}) as a system of ODEs, and we simulated the $N$ particles in MONDian gravity using the PM code, both with the same initial distribution. The two models are comparable because both must behave identical in the thermodynamic limit $N\longrightarrow\infty$. We also tested whether the distributions are stationary, as they should be for large $N$. In the external potential, the energies and angular momenta of the individual particles are strictly conserved, and so is the total energy and angular momentum. The single-particle kinetic energies and  momenta oscillate, and so does the total kinetic energy and the total momentum. We compare this with the exact system of Eqs.~(\ref{ddr})-(\ref{PDE}). Here, the total energy, total momentum, and total angular momentum are conserved. The PM code simulates this system. The fluctuations in the energies are shown in Fig.~\ref{Figure12}. Because the mesh in our PM approach breaks spherical symmetry, angular momentum also fluctuates slightly.


\section{Discussion}
\label{sect5}
\subsection{Accuracy of the method}
The inaccuracies in the algorithm arise from the fact that we used a fixed grid. The discrete grid has $n$ pixels in each dimension with a pixel width $s$ and has a finite volume $n^3s^3$.

The use of FFTs in the calculation of the potentials and acceleration fields generally gives rise to anti-aliasing: The functions obtain nonphysical short-wavelength oscillations. This effect is particularly strong when the point particles are represented with isolated nonzero pixel values in the mass density, but also when the CIC method is used. Smoothing with a minimum variance of one pixel width $s$ in a spherical domain of radius $4s$ needs to be used to overcome this problem. Although the discretization creates a minimum size, we reach a subpixel spatial resolution by using interpolation. We found that the retrieval of the forces on the particles with a Gaussian smoothing gives the most accurate results. To further limit the aliasing and increase accuracy, we did not use the numerical acceleration field, but calculated the potential, and we derived the forces instead using the finite-difference gradient of this scalar function.

\begin{figure}[t]
\centering
\includegraphics{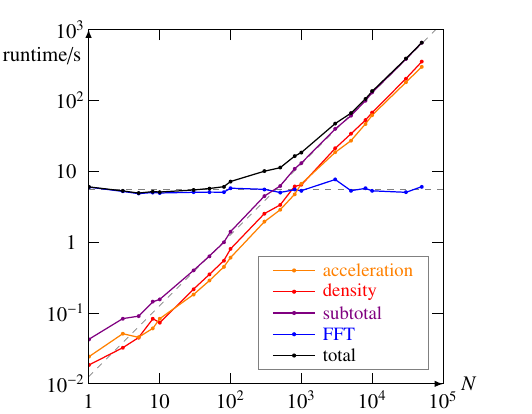}
\caption{Runtime vs.\ particle count $N$ for a $128^3$ grid with the Python code on an Intel Core i7-9750H. The time spend on the FFTs (blue) is roughly constant, and the time for the evaluation of the particle accelerations and/or densities grows linear with $N$ (orange, red, and purple). The total runtime shows a crossover around $10^3$ particles (black). The dashed lines are guides with slopes of zero and one.}
\label{Figure13}
\end{figure}

The runtime of our code has a time complexity of $O(Nn^3\log n)$, hence it scales linearly in the number of particles and almost linearly in the number of pixels because we made use of the FFT. The runtime as a function of particle count is plotted in Fig.~(\ref{Figure13}). For a $128^3$ grid, the evaluation of the accelerations or/and densities starts to dominate the runtime at $N=10^3$. Therefore, in the general case, we expect that the crossover occurs at a critical density of $10^3/128^3\approx 1/10^3$ particles per pixel.

\subsection{Improvements of the code}
The code could be made faster, allowing simulations with higher resolution/larger volumes, with smaller time steps/longer simulated times, or with more particles. We list below a few ideas for possible improvements that we did not implement.

\begin{inparaenum}[(i)]
\item 
The code is currently written in Python, using the FFTW library, which is a C library for the FFTs. The code could be entirely rewritten in a compiler programming language such as C, and compiled into fast-running machine code.

\item
A primitive data type can be chosen for the floats with half-precision, used by GPUs (with the cuFFTW library). The machine precision would be $10^{-3}$, which is close to the unavoidable error due to self-gravity. Its implementation should reduce memory use and speed up the code.

\item
Both the potential solver and the particle propagator in the code can be parallelized. The calculation of the potential requires $42$ FFTs ($3$ scalar plus $13$ vector FFTs). These three-dimensional FFTs need to be performed sequentially. For each direction, we have $n^2$ normal FFTs of a vector with $n$ elements. These can be performed in parallel. The increments of positions and velocities of the $N$ particles are independent, hence their calculations can also be made in parallel. Parallelizing the updating of the density is less straightforward. If the distance between two particles is smaller than 8 pixels, the density needs to be updated one particle at a time.

\item
If the density function changes only slightly at each time step, the potential solver needs just one iteration when it is seeded with the previous solution. This strategy can be employed when the spatial steps are smaller than the width of the weight functions (i.e.\ for small time steps), or when nearby particles have mostly  overlapping weight functions (i.e.\ for high number densities). Densities are also smooth when the matter is modeled as a fluid. We found that the PDE solver indeed sped up by a factor three.

\item
If changes in the fields at the time steps are small, we can calculate the increments $\Delta \rho_\text{B}$, $\Delta\boldsymbol g_\text{N}$, $\Delta\boldsymbol g_\text{M}$, $\Delta\boldsymbol F$, and $\Delta\boldsymbol H$, instead of the full fields. This will improve the accuracy and reduce the required number of significant digits so that half-precision can be used. For the nonlinear Eqs.~(\ref{inverse}) and (\ref{interpolation}), we can use the linearizations
\begin{align*}
\Delta\boldsymbol g_\text{M}(\boldsymbol r) &=
\nu\bigg(\frac{F}{a_0}\bigg) \Delta \boldsymbol F + \nu'\bigg(\frac{F}{a_0}\bigg) \frac{\boldsymbol F\bullet \Delta \boldsymbol F}{a_0 F} \boldsymbol F
, \\
\Delta \boldsymbol F(\boldsymbol r) &=
\mu\bigg(\frac{g_\text{M}}{a_0}\bigg) \Delta \boldsymbol g_\text{M} + \mu'\bigg(\frac{g_\text{M}}{a_0}\bigg) \frac{\boldsymbol g_\text{M}\bullet \Delta \boldsymbol g_\text{M}}{a_0 g_\text{M}} \boldsymbol g_\text{M}
.
\end{align*}
\end{inparaenum}

\subsection{Applications targeted to special systems}
We validated the PM code by simulating the two-body system and ring systems. For applications of systems with a high symmetry like this, however, the code could be adapted to make it more accurate and much faster. For example, a binary system can be simulated using ODEs and does not require a PDE solver, as shown in Sec.~\ref{sect4.1} for the deep MOND system Eq.~(\ref{system}). For the general case, the interpolation function $\mu$ is different, but the system is still governed by a two-body force that is a function of distance and the masses. One could now use the PDE solver of our code to calculate the force. This numerical force function can then be used in the ODE system for the actual simulations of the binary. 

In absence of an external field, the two-body system has a perfectly cylindrical symmetry: The potential is a function of the distance to and height on the axis of symmetry. By writing the PDE in cylindrical coordinates and using the Hankel transform, we limit the number of transforms in the PDE solver. The evaluation of the force function could thus be calculated fast, allowing one to test various choices of interpolation functions. For other systems with cylindrical symmetry, such as the motion of a test particle in a disk galaxy, this method can also be used to enhance the accuracy and speed \citep[see][for a study of rotation curves using the Hankel transform in this way]{Platschorre2019}.

The external field can be included in our PM code by fixing a plane at maximum distance from the particles in the cube where a permanent dipolar mass distribution is placed, that is, a layer of pixels with mass $M$ and an adjacent pixel layer of mass $-M$. When the external field is included, one can again evaluate the force function for the two-body problem numerically. Now, the force also depends on the polar angle (the angle between the direction of the external field and the vector $\boldsymbol r_2 -\boldsymbol r_1$), and not just on the interparticle distance.


\subsection{Extensions and future research directions}
Extensions to the code could make the code useful for more realistic modeling and other fields. Some examples are listed below.

\begin{inparaenum}[(i)]
\item Close encounters. If the ratio $Gm/s^2$ is much higher than $a_0$, the field on the pixel scale is Newtonian. It may be possible to obtain subpixel resolution for this case by calculating the near field with a particle-particle component.

\item Hydrodynamics. In galaxy clusters, the intergalactic gas component dominates the total mass contribution from the galaxies. One could implement Euler's equations for inviscid flow inside the MOND acceleration field. The evolution of the density $\rho_\text{B}$, pressure $p$, and flow field $\boldsymbol u$ can be calculated from the system
\begin{empheq}[left=\empheqlbrace]{align}
\dot{\rho}_\text{B} &= \boldsymbol u \bullet \mathscr{F}^{-1} \big( \mathrm i\boldsymbol k \widehat{\rho}_\text{B} \big) + \rho_\text{B} \mathscr{F}^{-1} \big( \mathrm i\boldsymbol k\bullet \widehat{\boldsymbol u} \big) , \nonumber \\
\dot{p} &= \boldsymbol u \bullet \mathscr{F}^{-1} \big( \mathrm i\boldsymbol k \widehat{p} \big) + \gamma p \mathscr{F}^{-1} \big( \mathrm i\boldsymbol k\bullet \widehat{\boldsymbol u} \big) , \nonumber \\
\dot{\boldsymbol u} &= \mathscr{F}^{-1} \big( \mathrm i \boldsymbol k \widehat{\cal E} \big) - \boldsymbol u \times \mathscr{F}^{-1} \big( \mathrm i\boldsymbol k \times \widehat{\boldsymbol u} \big) + \mathscr{F}^{-1} \big( \mathrm i\boldsymbol k \widehat{p} \big)/\rho_\text{B} + \boldsymbol g_\text{M}
. \nonumber
\end{empheq}
Here, $\widehat{\cal E} = \mathscr{F} \big( \boldsymbol u\bullet\boldsymbol u \big)/2$ is the transformed specific kinetic energy, and $\gamma$ is the polytropic constant. These are the Euler equations in Lamb form, where the derivatives are calculated using Fourier transforms. By running these on a GPU with cuFFTW in half-precision floats, one gains a speed-up of at least one order of magnitude.

\item Cosmology. In simulations of structure formation in the early Universe, one can introduce comoving coordinates to model the expansion of the Universe. This amounts to the equations for particle motion \citep{Peebles1993}
\[
\dot{\boldsymbol r}_i = \boldsymbol v_i/a , \quad
\dot{\boldsymbol v}_i = \boldsymbol g_\text{M}(\boldsymbol r_i) - \boldsymbol v_i \dot a/a ,
\]
and the density in Eq.(\ref{newtongrav}) can be replaced with $a\rho_\text{B}(\boldsymbol r)-a\overline{\rho_\text{B}}$, where the scale factor of the expanding Universe is given by $a(t)$.
\end{inparaenum}

\section{Conclusions}
\label{sect6}
We described, coded, and tested an algorithm for the simulation of the dynamical evolution of $N$ bodies for AQUAL-MOND. MOND is an alternative model for the gravity between stars, galaxies, and galaxy clusters, where standard theory requires CDM. We now list our final conclusions.

\begin{inparaenum}[(i)]
\item 
The MOND Eq.~(\ref{PDE}) is shown to be equivalent to the system Eqs.\ (\ref{newtongrav})-(\ref{interpolation}) of linear PDEs plus an algebraic equation. Therefore, linear PDE solvers can be used.

\item
Because we have a linear system, we can use Fourier transforms to solve it. Our PM code relies on $42$ FFTs, where a standard Poisson solver needs only two. The use of FFTs makes it faster than the finite-element codes N-MODY and RayMOND \citep{Londrillo2009,Candlish2014}, especially for large grids.

\item 
Four iterations are required to obtain an accuracy of $1\%$.

\item
Self-gravity is the dominant cause of the numerical error. We needed a Gaussian smoothing kernel of one pixel width over a spherical domain with a radius of four pixels. The standard CIC method used in Poisson solvers cannot be used.

\item
When the number of particles per pixel exceeds $1/10^3$, particle propagation takes more time than the PDE solver.

\item
We derived the analytic isothermal state Eqs.(\ref{phi})-(\ref{rho}) for the spherical deep MOND case. It is a two-parameter family, with a mass and a length scale. Because the deflection potential deviates significantly from that for a point mass, the solution could be relevant for elliptic galaxies and clusters.

\item 
We tested our PM code for systems where deep MOND analytic solutions \citep[by][]{Milgrom1994} are known and found good agreement. We also tested the isothermal state by creating a Maxwell-Boltzmann distribution with density Eq.~(\ref{rho}) and verified that the system is stationary.
\end{inparaenum}

Promising applications of the N-body MOND code include the numerical comparison with actual astrophysical observations on the Solar System, wide-binaries, and galaxy clusters.

\balance
\bibliographystyle{aa}
\bibliography{references}

\begin{appendix}
\section{Table of Symbols}
\label{AppZ}
\begin{table}[h]
\caption{List of symbols and notation.}
\label{table1}      
\centering
\begin{tabular}{ll}
\hline
\hline
symbol & quantity \\
\hline
$n$ & pixel count for each axis \\
$s$ & pixel width \\
$L=ns$ & system size \\ 
$N$ & particle count \\
$i=1\ldots N$ & particle index \\
\hline
$t$ & time \\
$\omega$ & orbital angular frequency \\
$\boldsymbol r=x\mathbf i+y\mathbf j+z\mathbf k$ & coordinate vector \\
$r$ & distance \\
$b$ & width of isothermal \\
$\boldsymbol k$ & wave vector \\
$k=|\boldsymbol k| $ & length of the vector \\
$m_i$ & mass of particle $i$ \\
$\boldsymbol r_i(t)$ & position of particle $i$ \\
$\boldsymbol v_i(t)=\dot{\boldsymbol r}_i$ & velocity of particle $i$ \\
$\boldsymbol g_i(t)=\dot{\boldsymbol v}_i$ & acceleration of particle $i$ \\
$\boldsymbol R(t)$ & center of mass \\
$\boldsymbol P$ & total momentum \\
$\boldsymbol L$ & total angular momentum \\
$E_\text{pot}$, $E_\text{kin}$ & potential, kinetic energy \\
$E=E_\text{pot}+E_\text{kin}$ & total energy \\
$M$ & total mass \\
\hline
$\phi(\boldsymbol r,t)$ & MOND potential \\
$\psi(x,y,t)$ & deflection potential \\
$\boldsymbol g_\text{M}(\boldsymbol r,t)=-\boldsymbol\nabla \phi$ & Milgromian gravity \\
$\boldsymbol g_\text{N}(\boldsymbol r,t)$ & Newtonian gravity \\
$\boldsymbol F(\boldsymbol r,t)=\boldsymbol g_\text{N}+\boldsymbol H$ & vector field \\
$\boldsymbol H(\boldsymbol r,t)$ & magnetic component of $\boldsymbol F$ \\
\hline
$c$ & speed of light \\
$a_0=1.2\, 10^{-10}\text{ms}^{-2}$ & Milgrom's constant \\
$G$ & Newton's constant \\
$\Lambda$ & cosmological constant \\
$a(t)$ & cosmological scale factor \\
$x=g_\text{M}/a_0$ & strength of MOND force \\
$y=F/a_0$ & strength of surface density \\
$\mu(x)=y/x$ & interpolation function \\
$\nu(y)=x/y$ & reciprocal interpolation \\
\hline
$\rho_\text{B}(\boldsymbol r,t)$ & baryonic matter density \\
$\rho_\text{D}(\boldsymbol r,t)$ & dark matter density \\
$\rho_\text{A}(\boldsymbol r,t)=\rho_\text{B}+\rho_\text{D}$ & apparent matter density \\
$\sigma(x,y,t)$ & surface mass density \\
\hline
$\Delta t$ & time step \\
$\mathscr{F}$ & 3d FT operator \\
$\wh{\boldsymbol f}(\boldsymbol k)=\mathscr{F}(\boldsymbol f)$ & 3d FT of $\boldsymbol f(\boldsymbol r)$ \\
$\mathbf i$, $\mathbf j$, $\mathbf k$ & standard basis vectors \\
$i$, $j$, $k$ & pixel coordinate index \\
$\xi$, $\eta$, $\zeta$ & random floating-points \\
\hline
\hline
\end{tabular}
\tablefoot{Symbol and significance of the physical quantities used. Vectors and vector fields are in boldface. The length of a vector or the strength of a vector field is in normal font. Most MOND research papers simply use $\boldsymbol g$ for the Milgromian gravity field $\boldsymbol g_\text{M}$.}
\end{table}

\section{Fourier transforms}
\label{AppA}
In the text, we did not distinguish much between continuous functions on $\mathbb R^3$ and the discrete representation on $(\mathbb Z/n\mathbb Z)^3$ for the $n\times n\times n$ grid. We simply considered the discrete functions as an approximation of the continuous functions on a finite interval. Let the physical dimension of the cubic pixels be $s$, and let the cubic volume have length $L=ns$. We assume that the functions smoothly drop to zero (or a constant) for $|x|>\tfrac{L}{2}$, $|y|>\tfrac{L}{2}$, $|z|>\tfrac{L}{2}$. The position coordinates that correspond to the grid points take on the values
\[
\boldsymbol r =
\begin{pmatrix} x \\ y \\ z \end{pmatrix} = 
\begin{pmatrix} i \\ j \\ k \end{pmatrix} s
, \quad
i, j, k \in \Big\{ {- \tfrac{n}{2}} , \ldots , \tfrac{n}{2} - 1 \Big\}
.
\]
Fourier space has the elements $\vec k$ as reciprocal vectors. It has the grid values
\[
\boldsymbol k =
\begin{pmatrix} k_1 \\ k_2 \\ k_3 \end{pmatrix} = 
\begin{pmatrix} i \\ j \\ k \end{pmatrix} \frac{2\pi}{L} , \quad
i, j, k \in \Big\{ {- \tfrac{n}{2}} , \ldots , \tfrac{n}{2} - 1 \Big\}
.
\]
The scalar function $\phi(\boldsymbol r)$ is expressed in its Fourier transform $\mathscr{F}(\phi)=\wh\phi(\boldsymbol k)$ via the expansion in plane waves, that is, the inverse Fourier transform
\[
\phi(\boldsymbol r) =
\frac{1}{(2\pi)^3}
\iiintop \wh{\phi}(\boldsymbol k) \mathrm e^{\mathrm i\boldsymbol k \bullet \boldsymbol r} \mathrm dk_1\mathrm dk_2\mathrm dk_3 \approx
\frac{1}{L^3}
\sum_{\boldsymbol k} \wh{\phi}(\boldsymbol k) \mathrm e^{\mathrm i\boldsymbol k \bullet \boldsymbol r}
,
\]
and the Fourier coefficients, that is, the direct Fourier transform is
\[
\wh{\phi}(\boldsymbol k) =
\iiintop \phi(\boldsymbol r) \mathrm e^{-\mathrm i\boldsymbol k \bullet \boldsymbol r} \mathrm dx\ \mathrm dy\ \mathrm dz \approx
s^3 \sum_{\boldsymbol r} \phi(\boldsymbol r) \mathrm e^{-\mathrm i\boldsymbol k \bullet \boldsymbol r}
.
\]

In our PM algorithm, we modeled point particles with a smooth density centered around the points $\boldsymbol r_i$. The standard CIC method uses the $n$-fold convolution of the box function $\text{box}(x/s)\text{box}(y/s)\text{box}(z/s)$ for the $n$th-order method \citep{Hockney1988}. Here, the box function is defined by
\[
\text{box}(y) =
\begin{cases}
1 & \text{if}\quad |y|<\tfrac{1}{2} , \\
0 & \text{else}
\end{cases}
.
\]
For large $n$, the convolution in one dimension and its Fourier transform may be approximated by
\begin{align*}
\frac{1}{s^n} \underbrace{\text{box}\bigg(\frac{x}{s}\bigg) * \cdots * \text{box}\bigg(\frac{x}{s}\bigg)}_\text{$n$ times} &\approx 
\frac{1}{s\sqrt{\pi n/6}} \exp\bigg(\frac{-6x^2}{ns^2}\bigg) , & \text{for} \quad n \gg 1
, \\
\bigg( \,\text{sinc}\frac{sk_1}{2} \bigg)^n &\approx \exp\bigg(\frac{-ns^2k_1^2}{24}\bigg) , & \text{for} \quad n \gg 1
.
\end{align*}
We therefore used the following density, with order $n=12$:
\[
\rho(\boldsymbol r) = \frac{m_i}{s^3(2\pi)^{3/2}} \text{exp}\bigg( \frac{-|\boldsymbol r-\boldsymbol r_i|^2}{2s^2} \bigg)
.
\]
The Fourier transform of this single-particle density is
\[
\wh{\rho}(\boldsymbol k) =
m_i
\text{exp}\bigg( \frac{-s^2|\boldsymbol k|^2}{2} \bigg) \mathrm e^{-\mathrm i\boldsymbol k\bullet \boldsymbol r_i}
.
\]
A standard deviation of a single pixel in position space implies a drop off of below $\mathrm e^{-\pi^2/2}=.007$ at the boundary of the reciprocal space. Furthermore, the density is very nearly spherically symmetric. The difference between the $12$th-order CIC function and the Gaussian with the same variance in three dimensions is $10^{-7}$ at most.

\newpage
\nobalance
\section{Flow scheme of the PM code}
\label{AppB}
This is the initialization procedure.
\begin{enumerate}
\item 
Create a lookup table for the exponential function.
\item 
Create a table of the $251$ coordinates in a sphere of radius $4$.
\item 
Create a $n^3$ matrix with the vector function $\boldsymbol k$.
\item 
Create a $n^3$ matrix for the inverse Laplace operator,
\[
\mathrm A = \begin{cases}
\dfrac{1}{\boldsymbol k\bullet\boldsymbol k} & \text{if}\quad  \boldsymbol k\neq\boldsymbol 0
, \\
0  & \text{if}\quad \boldsymbol k=\boldsymbol 0
.
\end{cases}
\]
\item Clear a $n^3$ matrix for the density: $\rho_\text{B}(\boldsymbol r) = 0$.
\item Loop through the particles $i=1\ldots N$.
\begin{itemize}
\item Set the mass $m_i$, and initialize the position $\boldsymbol r_i$ and velocity $\boldsymbol v_i$.
\item Next $i$.
End particle loop.
\end{itemize}
\end{enumerate}
This is the the main loop for time steps in the simulation.
\begin{enumerate}
\item Loop through all particles $i=1\ldots N$ (in case $n^3>251N$).
\begin{itemize}
\item Remove particle $i$ from the mass density by filling the pixels with zeros according to the list of $251$ pixels.
\item Next $i$.
End particle loop.
\end{itemize}
\item Loop through all particles $i=1\ldots N$.
\begin{itemize}
\item Implement the position step of Leapfrog: $\boldsymbol r_i \longrightarrow \boldsymbol r_i + \boldsymbol v_i \Delta t$.
\item Add particle $i$ to the mass density, with the list of $251$ pixels and the lookup table:
$\rho_\text{B}(\boldsymbol r) \longrightarrow \rho_\text{B}(\boldsymbol r) + \rho_i(\boldsymbol r)$, using Eq.~(\ref{smooth}).
\item Next $i$.
End particle loop.
\end{itemize}
\item FFT the density: $\wh\rho_\text{B}(\boldsymbol k) = \mathscr{F} (\rho_\text{B})$.
\item Calculate $\wh{\phi}_\text{N}(\boldsymbol k) = 4\pi G \mathrm A \wh\rho_\text{B}$.
\item Inverse FFT the potential: $\phi_\text{N}(\boldsymbol r) = \mathscr{F}^{-1} (\wh{\phi}_\text{N})$.
\item Find Newton gravity $\boldsymbol g_\text{N}(\boldsymbol r)$ from $\phi_\text{N}$ with finite differences.
\item Initialize the mass flux field by $\boldsymbol F(\boldsymbol r) = \boldsymbol g_\text{N}$.
\item Start the iteration loop.
\begin{itemize}
\item Calculate the gravity field: 
$\boldsymbol g_\text{M}(\boldsymbol r) = \nu\big(\sqrt{\boldsymbol F\bullet \boldsymbol F}/a_0\big)\boldsymbol F$.
\item FFT the gravity field: $\wh{\boldsymbol g}_\text{M}(\boldsymbol k) = \mathscr{F} (\boldsymbol g_\text{M})$.
\item Calculate the potential in Fourier space: $\wh\phi(\boldsymbol k) = \mathrm i \mathrm A \boldsymbol k\bullet \wh{\boldsymbol g}_\text{M}$.
\item Make the field conservative: $\wh{\boldsymbol g}_\text{M}(\boldsymbol k) = -\mathrm i\boldsymbol k \wh\phi$.
\item Exit the iteration loop at the fourth iteration.
\item Inverse FFT the gravity field: $\boldsymbol g_\text{M}(\boldsymbol r) = \mathscr{F}^{-1} (\wh{\boldsymbol g}_\text{M})$.
\item Calculate the field: $\boldsymbol F(\boldsymbol r) = \mu\big(\sqrt{\boldsymbol g_\text{M}\bullet \boldsymbol g_\text{M}}/a_0\big)\boldsymbol g_\text{M}$ .
\item Calculate the magnetic component, $\boldsymbol H(\boldsymbol r) = \boldsymbol F - \boldsymbol g_\text{N}$.
\item FFT the magnetic field: $\wh{\boldsymbol H}(\boldsymbol k) = \mathscr{F} (\boldsymbol H)$.
\item Make the field divergence free: $\wh{\boldsymbol H}(\boldsymbol k) \longrightarrow \wh{\boldsymbol H} - \boldsymbol k \mathrm A \boldsymbol k\bullet \wh{\boldsymbol H}$.
\item Inverse FFT the magnetic field: $\boldsymbol H(\boldsymbol r) = \mathscr{F}^{-1} (\wh{\boldsymbol H})$.
\item Calculate the field with $\boldsymbol F(\boldsymbol r) = \boldsymbol g_\text{N} + \boldsymbol H$.
\item Next iteration.
\end{itemize}
\item Inverse FFT the potential: $\phi(\boldsymbol r) = \mathscr{F}^{-1} (\wh\phi)$.
\item Loop through all particles $i=1\ldots N$.
\begin{itemize}
\item Evaluate $\boldsymbol g_i$, with the list of $251$ pixels and the lookup table using Eq.~(\ref{gi}).
\item Implement the velocity step of Leapfrog: $\boldsymbol v_i \longrightarrow \boldsymbol v_i + \boldsymbol g_i \Delta t$.
\item Next $i$.
End particle loop.
\end{itemize}
\item Update time $t\longrightarrow t + \Delta t$, and
\item Go to the next time step.
\end{enumerate}
\end{appendix}
\end{document}